\documentclass[aps,pra,twocolumn]{revtex4-1}
\usepackage{amsmath}
\usepackage{amsfonts}
\usepackage{graphicx}
\usepackage{color}
\usepackage{dsfont}
\usepackage{verbatim}
\usepackage{mathtools}
\usepackage[normalem]{ulem}
\usepackage{comment}
\usepackage{stackrel}
\usepackage{bm}
\usepackage[pdftex]{hyperref}
\hypersetup{colorlinks = true, urlcolor = black, linkcolor = black, citecolor = black}

\DeclareMathOperator{\sgn}{sgn}

\definecolor{myblue}{rgb}{.93, .93, 1}

\newcommand{\bsub}{\begin{subequations}}
	\newcommand{\esub}{\end{subequations}}

\newcommand{\vex}[1]{\bm{\mathrm{#1}}}

\begin{document}
	
	\title{Hofstadter butterfly and Floquet topological insulators in minimally twisted bilayer graphene}

\author{Yang-Zhi~Chou}\email{yzchou@umd.edu}
\affiliation{Condensed Matter Theory Center and Joint Quantum Institute, Department of Physics, University of Maryland, College Park, Maryland 20742, USA}

\author{Fengcheng~Wu} \email{wufcheng@umd.edu}
\affiliation{Condensed Matter Theory Center and Joint Quantum Institute, Department of Physics, University of Maryland, College Park, Maryland 20742, USA}

\author{Sankar Das Sarma}
\affiliation{Condensed Matter Theory Center and Joint Quantum Institute, Department of Physics, University of Maryland, College Park, Maryland 20742, USA}
		
\date{\today}

\begin{abstract}
We theoretically study the Hofstadter butterfly of a triangular network model in minimally twisted bilayer graphene (mTBLG). The band structure manifests periodicity in energy, mimicking that of Floquet systems.
The butterfly diagrams provide fingerprints of the model parameters and reveal the hidden band topology. In a strong magnetic field, we establish that mTBLG realizes low-energy Floquet topological insulators (FTIs) carrying zero Chern number, while hosting chiral edge states in bulk gaps. 
We identify the FTIs by analyzing the nontrivial spectral flow in the Hofstadter butterfly, and by explicitly computing the chiral edge states. 
Our theory paves the way for an {\it effective} practical realization of FTIs in equilibrium solid state systems.
\end{abstract}

\maketitle

\section{Introduction} 
Orientation misalignment in twisted bilayer graphene gives rise to a moir\'e superlattice greatly changing the electronic band structure. When the twist angle $\theta$ is around the magic angle ($\sim 1.1^{\circ}$), the low-energy moir\'e bands become nearly flat \cite{Bistritzer}, and strongly enhanced many-body interactions induce superconducting and correlated insulating states \cite{tbg1,tbg2,Lu2019,Polshyn2019,Yankowitz2019,Cao2019_electric,Jiang2019,Xie2019_spectroscopic,Choi2019,Codecido2019}. Away from the magic angle, the system can behave quite differently. The focus of this work is on {\it minimally} twisted bilayer graphene (mTBLG) with $\theta \ll 1^{\circ}$.
Although the flatband physics may not be significant here, mTBLG is a highly interesting system with intriguing properties worthy of serious experimental attention.

\begin{figure}[t]
	\includegraphics[width=0.425\textwidth]{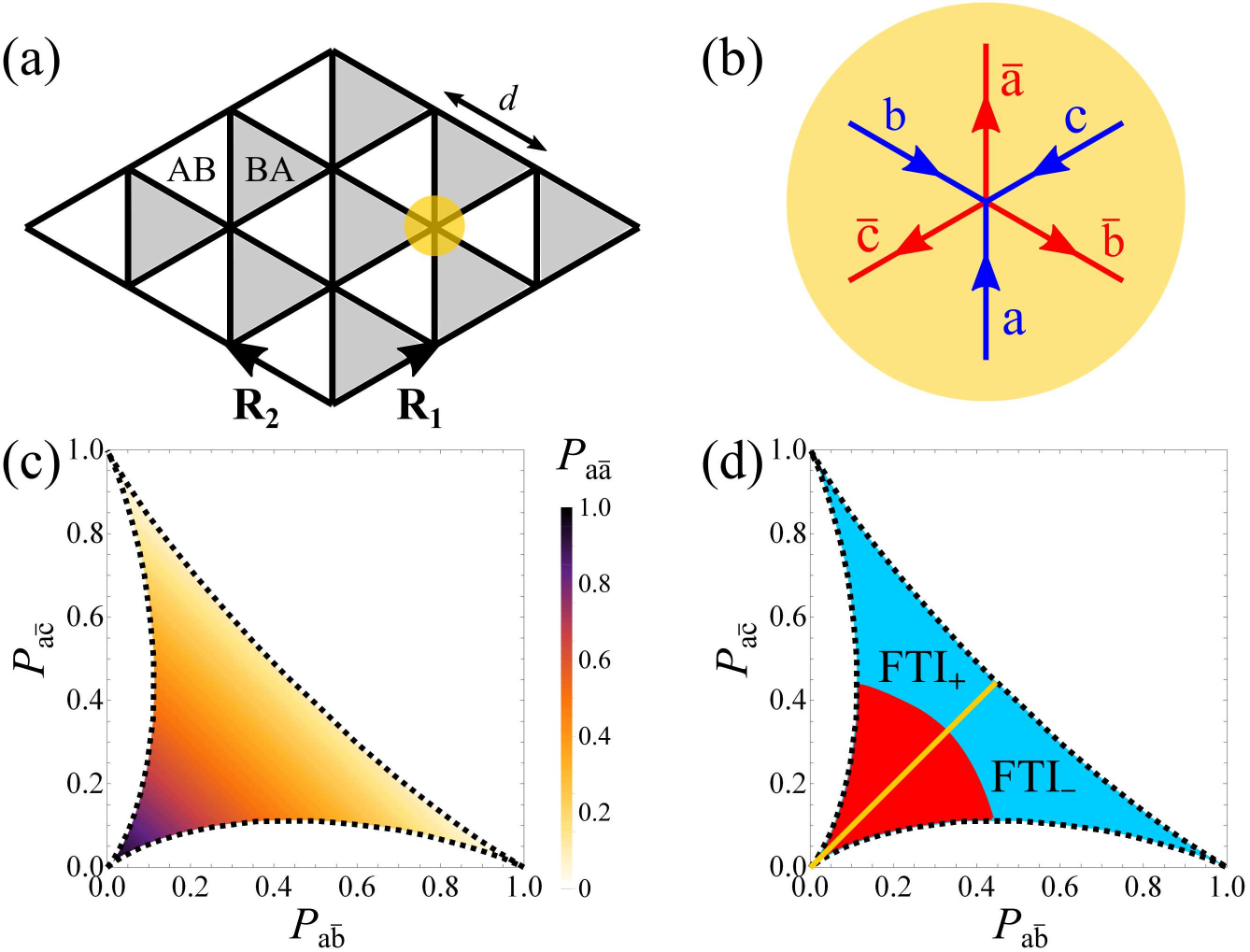}
	\caption{(a) Overview of the network model in mTBLG. (b) Domain-wall junction. (c) $P_{a\bar{a}}$ as a function of $P_{a\bar{b}}$ and $P_{a\bar{c}}$.  Allowed parameter space are enclosed by black dashed lines. (d) Topological phase diagram of the network model. The blue region hosts FTI phase with topological invariant $\nu=\pm 1$ for FTI$_\pm$; the red regime presents metallic phase with gapped Dirac cones but no full bulk gap; the yellow line indicates the gapless Dirac cones. 
	}
	\label{Fig:Fig1}
\end{figure}

The mTBLG hosts a triangular network of one-dimensional (1D) conducting channels when a large electric field is externally applied in the $\hat{z}$ direction (out of plane)  \cite{SanJose2013,Yoo2019_reconstruction,Rickhaus2018,Xu2019_GiantOscillation,Efimkin2018,Walet2019,Hou2019,Tsim2020,DeBeule2020,Huang2018Helical_TBLGexp,Ramires2018,Fleischmann2019}.
Low-energy electronic states in AB and BA regions are gapped out by the electric field, but domain walls separating the two regions support gapless 1D channels, which are valley Hall kink states with valley-dependent chiralities \cite{Martin2008,Jung2011,Zhang2013,Vaezi2013}. The AA regions act as junctions that connect domain-wall states along different directions [Fig.~\ref{Fig:Fig1}(b)]. The low-energy electronic structure in this system can be captured by a triangular network model \cite{Efimkin2018}. While energetics of the network model has been studied, a full characterization of the system, including its topological properties, remain an interesting open question.

In this article, we study the electronic structure of mTBLG in a magnetic field $B \hat{z}$ by calculating the Hofstadter butterfly of the network model. In the low $B$ field regime, we show that Dirac points can reemerge at finite fields, which should provide fingerprints for determining the network model parameters. In the strong $B$ field regime, the probabilities of an electron to deflect right and left at a scattering junction can be significantly different, which leads to gap opening at Dirac points. In such a case, we demonstrate that the network model effectively realizes Floquet topological insulators
(FTIs) \cite{Kitagawa2010,Rudner2013,Liang2013}, where Chern numbers of bulk bands are zero, 
but there are chiral edge states traversing bulk gaps. In realistic mTBLG, the bands at the ultraviolet energies set by the gap at AB/BA region can have nonzero Chern numbers, but the low-energy spectrum is well described by the FTIs.
We identify the FTIs based on a nontrivial spectral flow in the Hofstadter butterfly as well as an explicit calculation of the chiral edge states. We also discuss experimental implications of our work, predicting the exciting possibility of realizing a solid state equilibrium FTI in mTBLG.

The rest of the article is organized as follows: We describe the network model and construct the scattering matrix in Sec.~\ref{Sec:Model}. The band structures of the network model are calculated in Sec.~\ref{Sec:Bands}. Then, we compute the Hofstadter butterfly in the weak $B$ field regime and show that the Dirac points can be used to identify the model parameters in Sec.~\ref{Sec:butterfly}. In Sec.~\ref{Sec:FTI}, we predict that the FTI phase in the strong $B$ field regime. The experimental signatures and open questions are discussed in Sec.~\ref{Sec:Discussion}. The detailed calculations for the Hofstadter butterfly are presented in Appendix~\ref{APP:B_field}. In Appendix~\ref{APP:Chern_N}, we compute the Berry curvature of the bands in the FTI phase. We explain the setup for calculating the chiral edge state of a FTI in Appendix~\ref{APP:OBC_cal}.

\section{Model}\label{Sec:Model}

In the network model of mTBLG [Fig.~\ref{Fig:Fig1}], each domain-wall link that connects adjacent junctions has a length $d$ (i.e., the moir\'e periodicity) and hosts two 1D channels per spin and per valley. We neglect the spin index thereafter, and treat the electronic structures in the two valleys K and K$'$ separately since the system is smooth at the atomic scale. Noting that the two valleys are related by $\mathcal{C}_{2z}$ symmetry (twofold rotation around  $\hat{z}$ axis), we mainly focus on states in K valley.  We further assume that the two 1D channels in K valley are decoupled at the domain wall and at the junctions, which leads to a single-channel network model \cite{Efimkin2018}.

In the single-channel model, an electron moves chirally along domain-wall links but experiences scattering at junctions [Fig.~\ref{Fig:Fig1}(b)] that can be formulated by $\hat{\Psi}_{\text{out}}(\vex{x}_0)=\hat{S}\hat{\Psi}_{\text{in}}(\vex{x}_0)$, where $\vex{x}_0$ is the position of the junction, $\hat{\Psi}_{\text{in}}=\left[\psi_a, \psi_b, \psi_c\right]^{\top}$ and $\hat{\Psi}_{\text{out}}=\left[\psi_{\bar{a}} , \psi_{\bar{b}}, \psi_{\bar{c}}\right]^{\top}$ represent the incoming and outgoing wavefunctions respectively, and $\hat{S}$ is a $3\times 3$ {\it unitary} scattering matrix. The form of $\hat{S}$ is dictated by symmetries, in particular, $\mathcal{C}_{3z}$ (threefold rotation around $\hat{z}$) and $\mathcal{C}_{2z} \mathcal{T}$ ($\mathcal{C}_{2z}$ combined with spinless time-reversal symmetry $\mathcal{T}$). Under the $\mathcal{C}_{3z}$ condition, $\hat{S}$ can be modeled as follows
\begin{align}\label{Eq:S_matrix_C3}
\hat{S}=\left[\begin{array}{ccc}
\alpha & \zeta  & \xi\\
\xi & \alpha & \zeta \\
\zeta & \xi & \alpha
\end{array}\right].
\end{align}
To satisfy the unitarity $\hat{S}^{\dagger} \hat{S} = \mathbb{I}$ , we use the following parameterization $\alpha=\sqrt{P_{a \bar{a}}}$, $\zeta=e^{-i\frac{2\Phi+\Phi'}{3}}\sqrt{P_{a \bar{c}}}$, $\xi=e^{-i\frac{\Phi+2\Phi'}{3}}\sqrt{P_{a \bar{b}}}$,
$\Phi=\cos^{-1}\left(\frac{\alpha^2|\xi|^2-|\zeta|^2|\xi|^2-\alpha^2|\zeta|^2}{2\alpha|\zeta|^2|\xi|}\right)$, and
$\Phi'=\cos^{-1}\left(\frac{\alpha^2|\zeta|^2-|\zeta|^2|\xi|^2-\alpha^2|\xi|^2}{2\alpha|\zeta||\xi|^2}\right)$. 
Here $P_{a \bar{a}}$ is the probability for forward scattering, while $P_{a \bar{b}}$ and $P_{a \bar{c}}$ are those for deflected scatterings. In addition to $P_{a \bar{a}}+P_{a \bar{b}}+P_{a \bar{c}}=1$, these probabilities are further constrained by the requirements that the phases $\Phi$ and $\Phi'$ are real valued [Fig.~\ref{Fig:Fig1}(c)]. We note that $\Phi$ and $\Phi'$ are defined modulo $2\pi$. Changing $\Phi$ and $\Phi'$ to $-\Phi$ and $-\Phi'$ respectively also satisfies the unitary condition, but the obtained energy spectrum changes by an overall minus sign.  There are in total 6 ways to construct $\hat{S}$ for a given $(P_{a\bar{a}},\gamma)$.
We make a particular choice without loss of generality. The findings and the conclusions in this work do not depend on this specific choice.

\begin{figure*}[t]
	\includegraphics[width=0.9\textwidth]{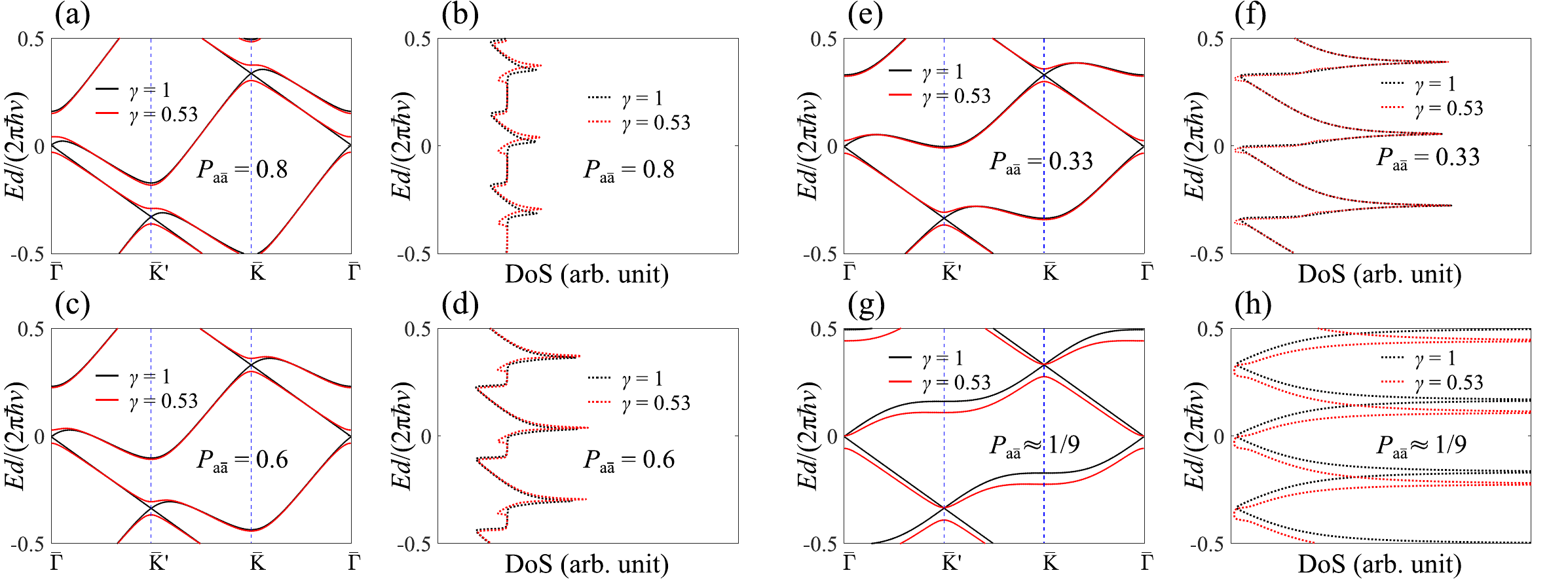}
	\caption{Band structure and density of states for $\gamma=1$ and $\gamma=0.53$ ($\gamma\equiv P_{a\bar{b}}/P_{a\bar{c}}$). (a), (c), (e) Band structures with lines connecting $\bar\Gamma$, $\bar{\text{K}}'$, and $\bar{\text{K}}$ points. (b), (d), (f) Density of states. All the energy levels are shifted such that the $\bar\Gamma$ Dirac nodes appear at zero energy. The density of states plots are in the same scale but in an arbitrary unit.}
	\label{Fig:Fig2}
\end{figure*}

The symmetry $\mathcal{C}_{2z} \mathcal{T}$ requires that $\hat{S}=\hat{S}^\top$,   and therefore, $P_{a \bar{b}}=P_{a \bar{c}}$. The $\hat{S}$ matrix in Eq.~(\ref{Eq:S_matrix_C3}) is equivalent to that in Ref.~\cite{Efimkin2018} when $P_{a \bar{b}}=P_{a \bar{c}}$, but describes a generalization allowing a broader range of parameters including $P_{a\bar{b}}\neq P_{a\bar{c}}$.

\begin{figure}[t]
	\includegraphics[width=0.425\textwidth]{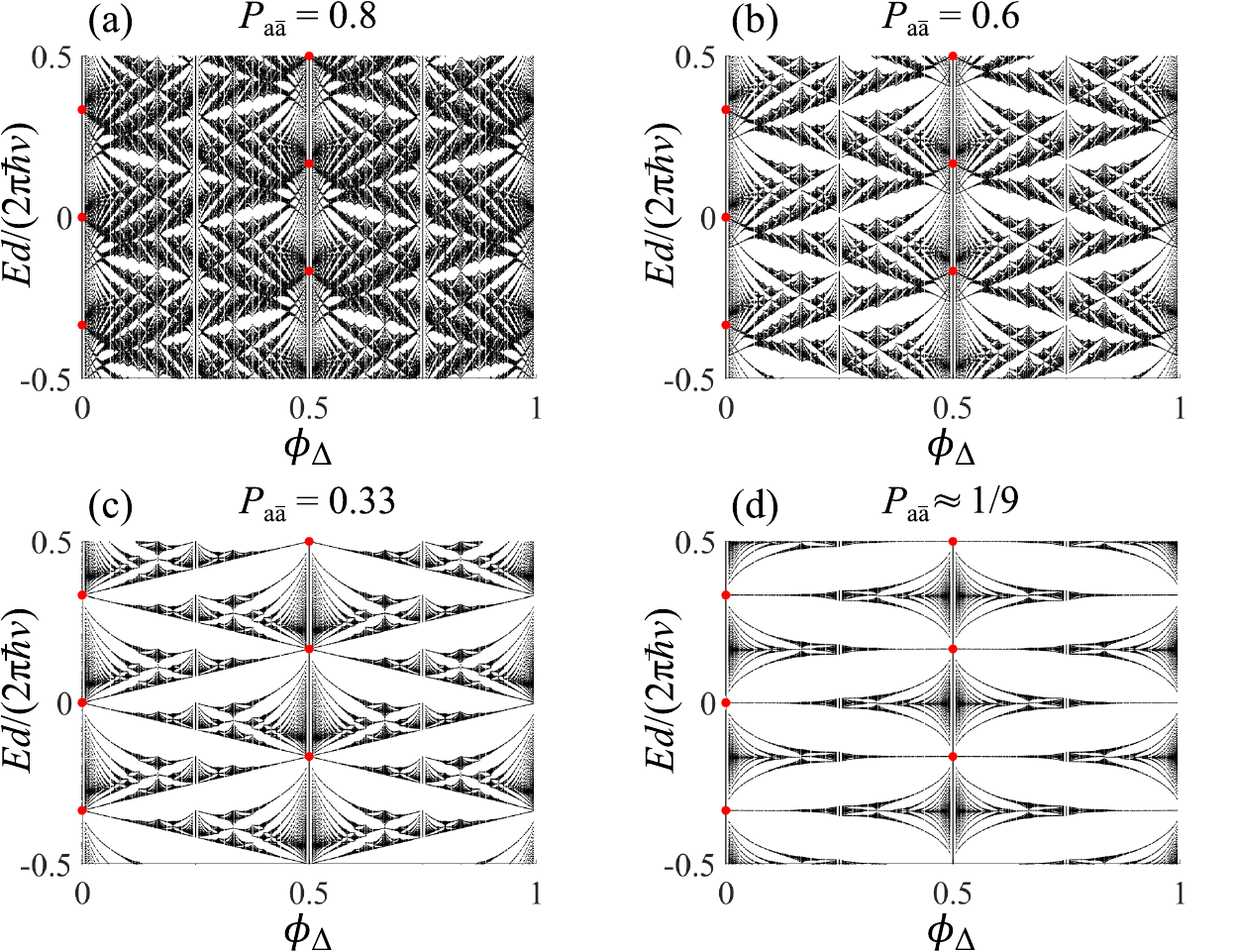}
	\caption{Hofstadter butterfly of the network model with $\gamma\equiv P_{a\bar{b}}/P_{a\bar{c}}=1$. Red dots mark Dirac points at $\phi_{\triangle}=0$ and $1/2$. 
	}
	\label{Fig:Fig3}
\end{figure}

\section{Band structure}\label{Sec:Bands}

The  network model in Fig.~\ref{Fig:Fig1} describes a periodic array of junctions in a triangular lattice with lattice vectors $\vex{R}_1=(\sqrt{3}d/2,d/2)$ and $\vex{R}_2=(-\sqrt{3}d/2,d/2)$. The electronic structure can be computed by combining the Bloch's theorem and the scattering matrix $\hat{S}$ as follows
\begin{align}\label{Eq:Translation}
&\left[\!\!\begin{array}{c}
\psi_{\bar a}(\vex{x}_0+\vex{R}_1+
\vex{R}_2)\\
\psi_{\bar b}(\vex{x}_0-
\vex{R}_2)\\
\psi_{\bar c}(\vex{x}_0-\vex{R}_1)
\end{array}\!\!
\right]\!
=\hat{T}\hat\Psi_{\text{in}}(\vex{x}_0)=e^{i\frac{E d}{\hbar v}} \hat\Psi_{\text{out}}(\vex{x}_0),\\
\label{Eq:band}&\rightarrow\left(e^{-i\frac{Ed}{\hbar v}}-\hat{T}^{-1}\hat{S}
\right)\hat\Psi_{\text{in}}(\vex{x}_0)=0,
\end{align}
where $\hat{T}=\text{diag}\left[e^{i\vex{k}\cdot\left(\vex{R}_1+\vex{R}_2\right)},e^{-i\vex{k}\cdot\vex{R}_2},e^{-i\vex{k}\cdot\vex{R}_1}\right]$ is the translation operator with $\vex{k}$ being the Bloch wave vector, $e^{i\frac{Ed}{\hbar v}}$ describes the phase accumulation in traveling along a link, $E$ is the energy, and $v$ is the domain-wall state velocity. To derive Eq.~(\ref{Eq:band}), we have replaced $\hat{\Psi}_{\text{out}}(\vex{x}_0)$ in Eq.~(\ref{Eq:Translation}) by $\hat{S}\hat{\Psi}_{\text{in}}(\vex{x}_0)$.

The energy spectrum $E(\vex{k})$ is simply given by the phase of the eigenvalues of $\hat{T}^{-1}\hat{S}$. Thus, $E(\vex{k})$ has an infinitely repeated pattern with a period $2\pi \hbar v/d$ \cite{Efimkin2018}  due to the resemblance of Eq.~\eqref{Eq:band} to the Floquet problem \cite{Pasek2014,Potter2020,Kim2020}. The occurrence of infinitely many bands is a consequence of the continuous linear dispersion approximation in each 1D domain-wall state. 
In realistic mTBLG systems, we expect a large but finite number of repeated bands within the energy gap opened in AB/BA regions.

We plot the band structures and the density of states in Fig.~\ref{Fig:Fig2} for representative values of $P_{a \bar{a}}$ and $\gamma\equiv P_{a\bar{b}}/P_{a\bar{c}}$. 
We first focus on a few special limits as follows: 
\begin{enumerate}
	\item $P_{a\bar{a}}=1$: This limit realizes a system with decoupled chiral chains along three orientations of the domain-wall links, $\hat{y}$, $\frac{\sqrt{3}}{2}\hat{x}-\frac{1}{2}\hat{y}$, and $-\frac{\sqrt{3}}{2}\hat{x}-\frac{1}{2}\hat{y}$. The band structure is simply a linear superposition of three chiral dispersions and is insensitive to an external magnetic field.

	\item $P_{a\bar{a}}=1/9$ and $P_{a\bar{b}}=P_{a\bar{c}}=4/9$: The band structure contains nearly particle-hole symmetric Dirac nodes. The low energy properties near the Dirac points agree well with the Dirac equation, e.g., the Landau level spectrum.
	
	\item $P_{a\bar{b}}=1$ ($P_{a\bar{c}}=1$): The electrons move clockwise (counterclockwise) around localized isolated triangular loops for $P_{a\bar{b}}=1$ ($P_{a\bar{c}}=1$). The band structure features three flat bands in an energy period $2\pi\hbar v/d$. The energies of those flat bands are $E=\pm\frac{1}{3}\pi\hbar v/d,\pi\hbar v/d$ within the first energy period for an integer magnetic flux per triangle.
	
\end{enumerate}

In the presence of $\mathcal{C}_{3z}$ and $\mathcal{C}_{2z} \mathcal{T}$ symmetries ($\gamma=1$), the band structures show generic features that are independent of $P_{a \bar{a}}$ value: (i) there are three bands within one energy period $2\pi \hbar v/d$; (ii) each pair of adjacent bands are connected via a Dirac cone at one of the three high-symmetry momenta $(\bar{\Gamma}, \bar{\text{K}}, \bar{\text{K}}')$ in the moir\'e Brillouin zone; (iii) Dirac velocity is always $v/2$, precisely half of the domain-wall state velocity \cite{Efimkin2018}. However, details of the bands do vary with $P_{a \bar{a}}$. Particularly, particle-hole asymmetry of the Dirac cones becomes stronger as $P_{a \bar{a}}$ approaches 1 (decoupled 1D chiral chains limit).

In the absence of $\mathcal{C}_{2z}\mathcal{T}$ ($\gamma\neq 1$), direct band gaps open at the Dirac nodes, but the system might not develop a full band gap due to the complications of electron pockets. Our theory for both $\gamma=1$ and $\gamma\neq 1$ can be generalized to a two-channel network model, and we expect that our qualitative predictions still apply.

\section{Hofstadter butterfly and Dirac points}\label{Sec:butterfly}

In the presence of a magnetic field $B\hat{z}$, the phase accumulation of an electron traveling along a link $\ell$ becomes $\exp(i \frac{Ed}{\hbar v}-i \frac{e}{\hbar} \int_\ell \vex{A} \cdot d \vex{r} )$, where $-e<0$ is the electron charge and $\vex{A}$ is the vector potential.
%This model is strictly valid as long as the magnetic length ($l_B=1/\sqrt{eB}$) is much larger than $d$.
In the Landau gauge $\vex{A}=(0,Bx)$, the system is translationally invariant in $\hat{y}$ direction but generically not along $\hat{x}$. We focus on the \textit{commensurate} cases such that $eB\sqrt{3}d^2/ h=p/q$ where $p$ and $q$ are integers. The flux per triangle area ($\mathcal{A}_{\triangle}=\sqrt{3}d^2/4$) in unit of the flux quantum $h/e$ is thus
\begin{align}
\phi_{\triangle}=\frac{p}{4q}.
\end{align}
Under such a flux, we choose a magnetic unit cell with lattice vectors $\vex{\mathcal{R}}_1=(q\sqrt{3}d,0)$ and $\vex{\mathcal{R}}_2=(0,d)$. We calculate the Hofstadter butterfly (i.e., the energy spectra as a function of $\phi_{\triangle}$) by choosing $q$ to be prime numbers $\le 37$ and $p=0,1,2,..,4q-1$. In Appendix~\ref{APP:B_field}, we discuss how to compute the spectrum in the presence of a magnetic field.

\begin{figure}[t!]
	\includegraphics[width=0.425\textwidth]{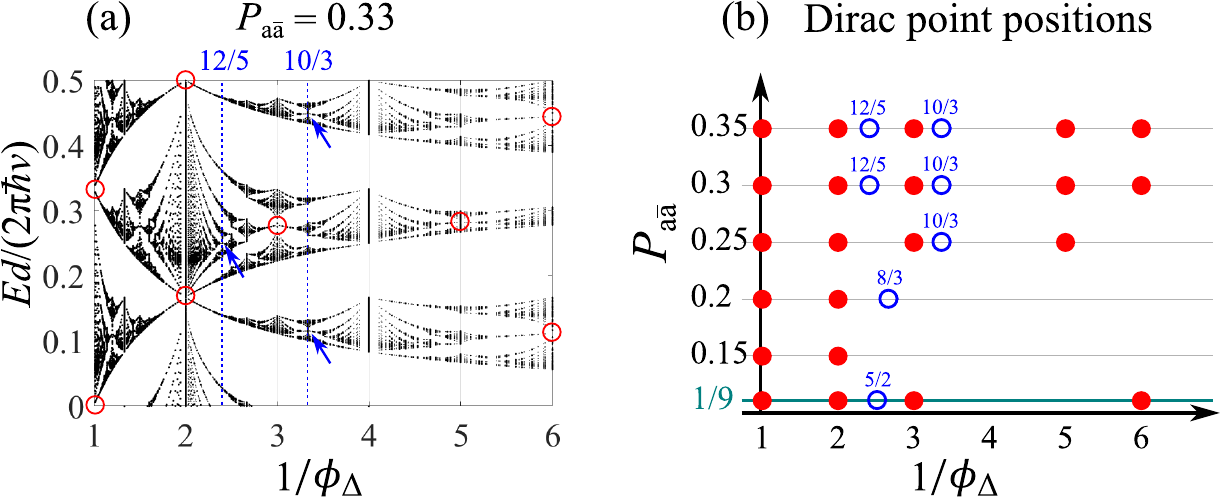}
	\caption{(a) Dirac points at a weak magnetic field. We plot the energy levels as a function of $1/\phi_{\triangle}$ such that the Dirac points are clearly seen.
	Red circles (blue arrows) mark the Dirac points at integer (fractional) values of $1/\phi_{\triangle}$.
	(b) A summary of the Dirac point positions for different values of $P_{a\bar{a}}$. Red solid dots represent the Dirac points at the integer values; blue circles indicate those at a few numerically resolvable fractional values. 
	Dirac points for $1/\phi_{\triangle} \in (1,2)$ are not explicitly shown and can be identified using the spectral symmetry between $1-\phi_{\triangle}$ and $\phi_{\triangle}$.
	}
	\label{Fig:Fig4}
\end{figure}

Another effect of the $B$ field is to modify the $\hat{S}$ matrix, since $\mathcal{C}_{2z} \mathcal{T}$ symmetry is explicitly broken by the $B$ field. Therefore, the ratio $\gamma \equiv P_{a\bar{b}}/P_{a\bar{c}}$ can deviate from 1. Microscopically, the magnetic field shifts the domain-wall states away from the zero-field positions \cite{Li2016_nanotechnology,Wang2017,Winn2019}, and the junction is subsequently modified \cite{Li2018}. 
We first consider the case where the $B$ field is small and $\gamma$ can be approximated by its zero-magnetic-field value (i.e., $\gamma=1$). 
In Fig.~\ref{Fig:Fig3}, we plot the Hofstadter butterfly for $\phi_{\triangle} \in [0, 1]$  in one energy period ($2\pi \hbar v/d$) for different values of $P_{a\bar{a}}$ with $\gamma=1$. 
Independent of $P_{a\bar{a}}$, the electronic structure develops fractal energy gaps for generic $\phi_{\triangle}$ except for $\phi_{\triangle}=1/2$, where the bands are similar to that at zero field and also feature Dirac cones with velocity $v/2$. In addition, the energy spectra of $1-\phi_{\triangle}$ and $\phi_{\triangle}$ are identical given that $\gamma=1$.

Details of the Hofstadter butterfly depend on the value of $P_{a\bar{a}}$.
We first focus on $\phi_{\triangle} \ll 1$ where Landau levels (LLs) emanating from zero-field Dirac cones can be identified.
In particular, for $P_{a\bar{a}}=1/9$, the energy of LLs can be accurately fitted by $E=\pm (v/2) \sqrt{2\hbar e B n}$ for $n=0, 1, 2 ...$, as expected from the LL spectrum of 2D massless Dirac fermions. For other $P_{a\bar{a}}$ values, the LLs are complicated because of particle-hole asymmetry and other electron pockets coexisting with the Dirac cones in the zero-field band structure [Fig.~\ref{Fig:Fig2}].
It is worth mentioning that the spectrum has no $\phi_{\triangle}$ dependence when $P_{a\bar{a}}=1$ (decoupled 1D chiral chains limit). 
Going beyond the low field limit, Dirac cones reemerge in the spectra at certain values of $\phi_{\triangle}$. In Fig.~\ref{Fig:Fig4}(a), we plot the energy levels versus the \textit{inverse} flux $1/\phi_{\triangle}$.
The occurrence of the Dirac nodes sensitively depends on the value of $P_{a\bar{a}}$ as we show in Fig.~\ref{Fig:Fig4}(b), which could be used to estimate the $P_{a\bar{a}}$ value from transport experiments.

\begin{figure}[t!]
	\includegraphics[width=0.425\textwidth]{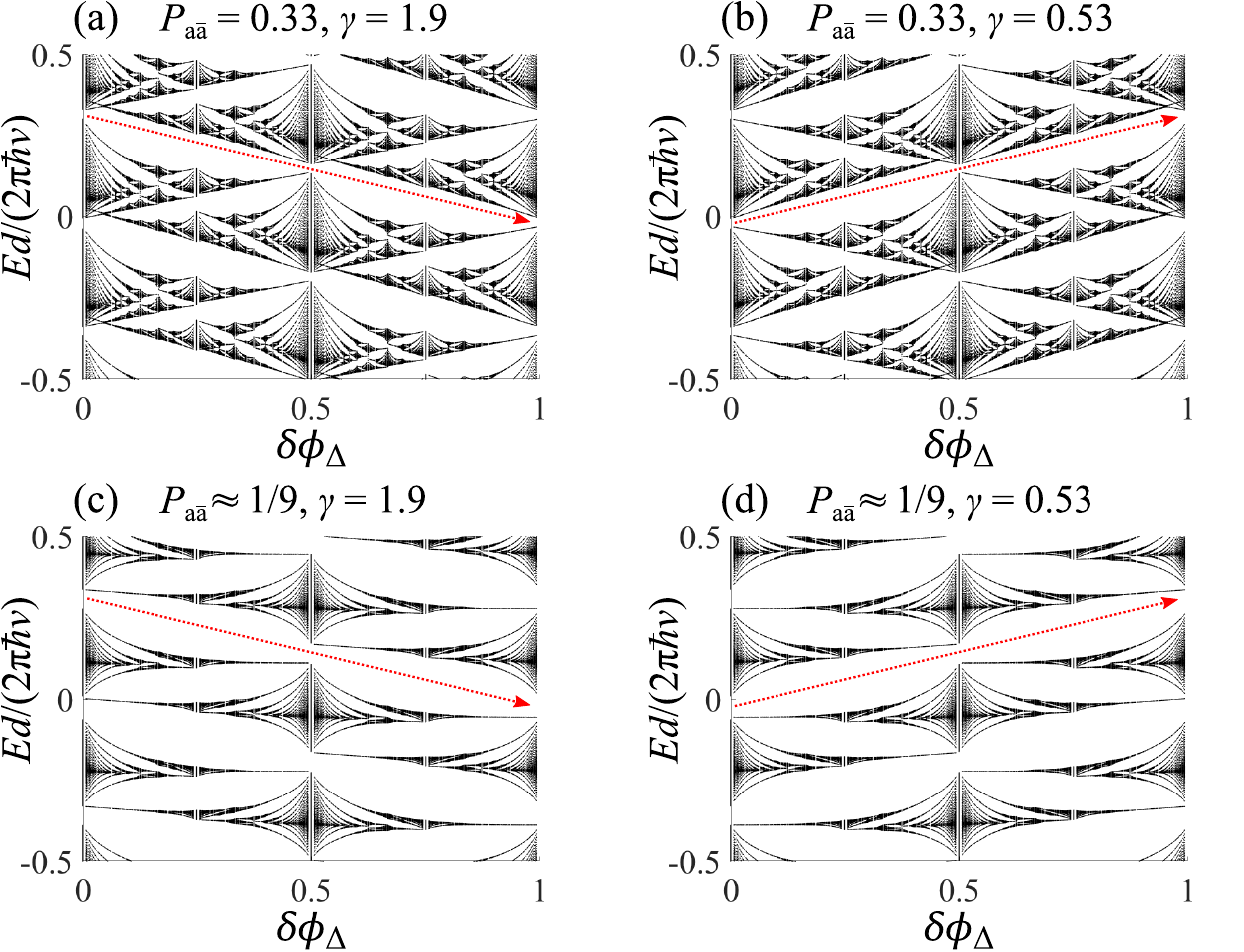}
	\caption{Hofstadter butterfly for network model with $\gamma\neq 1$. 
	The red dotted arrows indicate the spectral flow, which implies nontrivial band topology. $\delta \phi_{\triangle}$ is $\phi_{\triangle}-N$, where $N \geq 1$ is an integer.}
	\label{Fig:Fig5}
\end{figure}

\section{Floquet topological insulators}\label{Sec:FTI}

We now turn to the case where the $B$ field is strong enough such that $\gamma$ has a noticeable deviation from 1. 
For definiteness, we consider a flux $\phi_{\triangle} \in [N, N+1]$, where $N$ represents a large integer. Within this field range, we assume that $P_{a \bar{a}}$ and $\gamma \neq 1$ do not vary with $B$. The corresponding Hofstadter butterfly is shown in Fig.~\ref{Fig:Fig5} for representative values of $(P_{a \bar{a}}, \gamma)$.

\begin{figure}[t!]
	\includegraphics[width=0.425\textwidth]{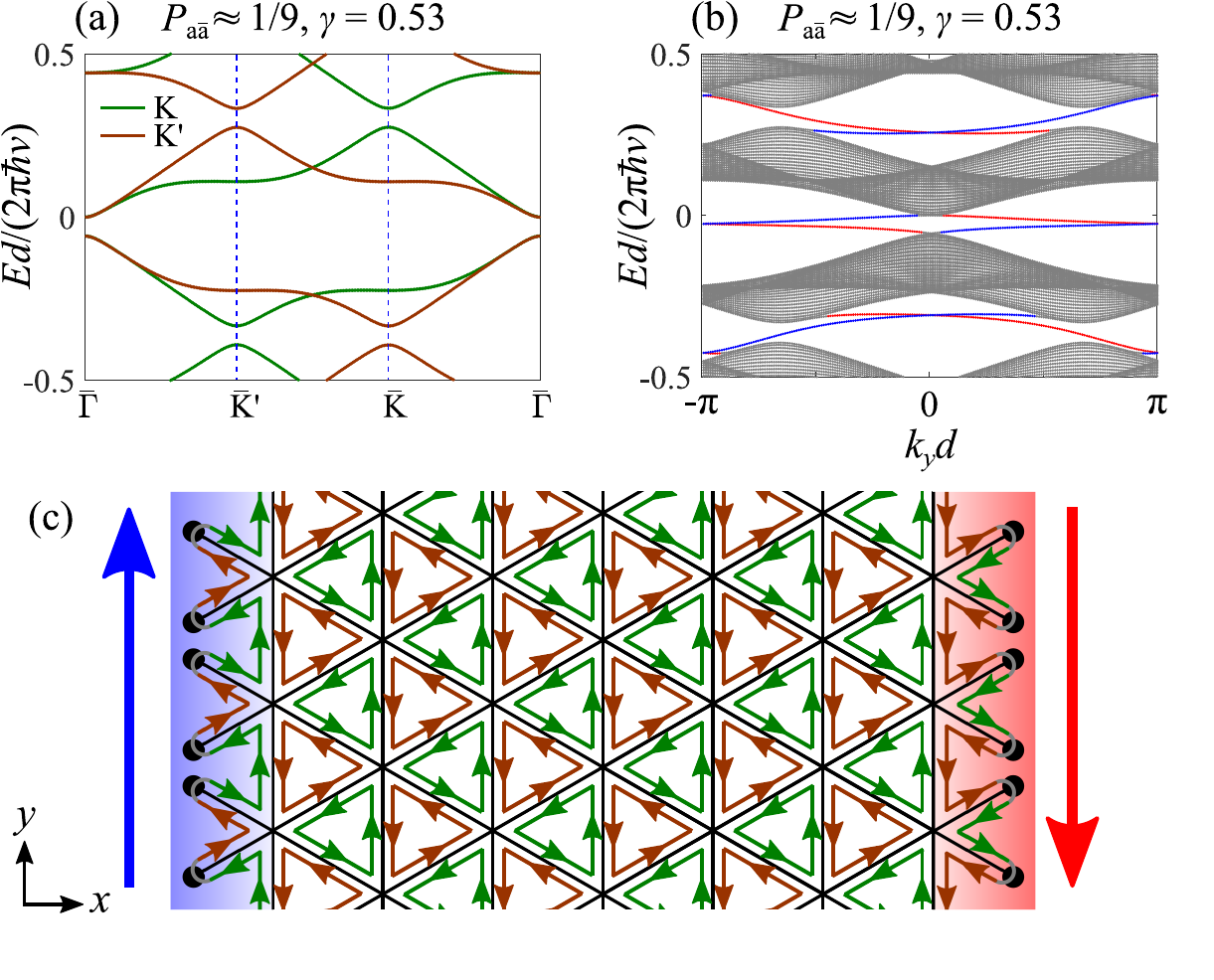}
	\caption{(a) The bulk band structures of K (green) and K$'$ (brown) valleys for $P_{a\bar{a}}=1/9$ and $\gamma=0.53$ . (b) Similar as (a) but for the band structure in a stripe geometry. The blue (red) lines represent chiral states localized on the left (right) edge. $\phi_{\triangle}=N \geq 1$ is assumed in (a) and (b). (c) Illustration of the stripe geometry and the chiral edge states as skipping orbits in the $P_{a\bar{c}}=1$ limit. The green (brown) arrows indicate electron motion in the K (K$'$) valley.
	}
	\label{Fig:Fig6}
\end{figure}

Dirac cones at $\phi_{\triangle}=N$ are gapped out because $\gamma \neq 1$, as demonstrated in Fig.~\ref{Fig:Fig5}. We focus on the parameter space where this Dirac gap opening leads to a full spectral gap, i.e., the blue regions in Fig.~\ref{Fig:Fig1}(d).
Physically, an observable gap needs to exceed thermal or disorder broadening---explaining why a strong magnetic field is required in practice.

While the spectra are identical at $\phi_{\triangle}=N$ and $N+1$ under the approximation that $\gamma$ is fixed, there is a nontrivial spectral flow from  $\phi_{\triangle}=N$ to $N+1$, which signals a nontrivial band topology \cite{Asboth2017}. In particular, the $n_0$th gap at $\phi_{\triangle}=N$ evolves to the $n_1$th gap at $\phi_{\triangle}=N+1$, where $n_1$ is $n_0-1$ and $n_0+1$, respectively, for $\gamma>1$ and $\gamma<1$. We note that the difference $\nu \equiv n_1-n_0$ is well defined, although $n_0$ and $n_1$ individually are not because the spectrum is \textit{unbounded} in our continuum approximation. The spectral flow can be obtained analytically in the $P_{a\bar{b}}=1$ ($\gamma\rightarrow \infty$) or the $P_{a\bar{c}}=1$ ($\gamma=0$) limit. For $P_{a\bar{b}}=1$ ($P_{a\bar{c}}=1$), electrons move clockwise (counterclockwise) in {\it localized} chiral loops encircling isolated triangles [Fig.~\ref{Fig:Fig6}(c)], and the entire spectrum is linearly shifted by $-1/3$ $\left(1/3\right)$ of $2\pi \hbar v/d$ with the increase of one unit flux per triangle [see Fig.~\ref{Fig:S3} and discussions in Appendix~\ref{APP:B_field}], which precisely leads to $\nu=-1$ ($\nu=1$). Here $\nu$ serves as a topological invariant that counts $n_{\text{edge}}$, the number of chiral edge modes, as proved in Ref.~\cite{Asboth2017}. Remarkably, the invariant $\nu$ takes the same value for every gap at $\phi_{\triangle}=N$. This implies that the Chern number for each band at $\phi_{\triangle}=N$ is zero, which we confirm through an explicit calculation as described in the Appendix~\ref{APP:Chern_N}.   

\begin{figure}[h]
	\includegraphics[width=0.45\textwidth]{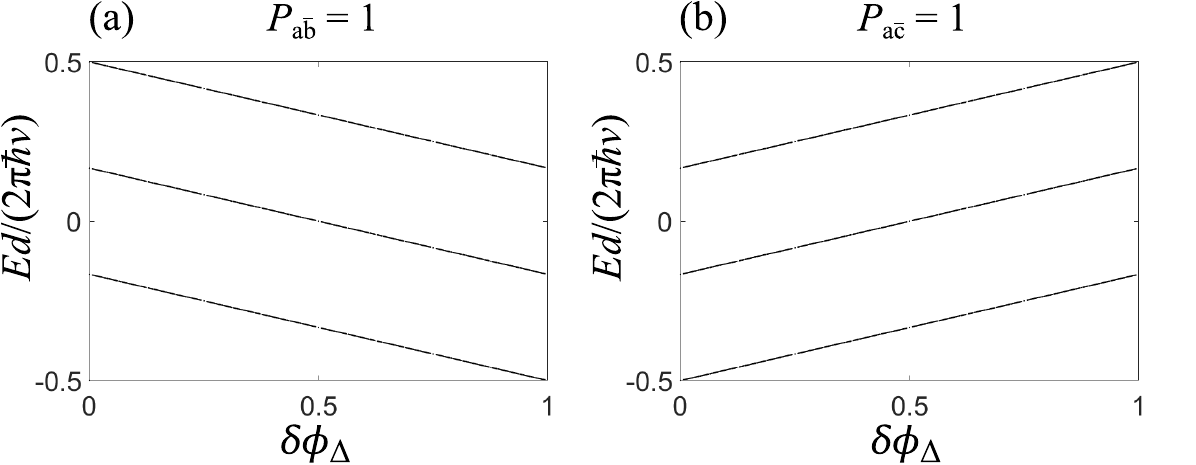}
	\caption{Spectral flows in (a) $P_{a\bar{b}}=1$ and (b) $P_{a\bar{c}}=1$ limit.}
	\label{Fig:S3}
\end{figure}

We establish the existence of chiral edge states by calculating the band structure on a stripe with an open (periodic) boundary condition along $\hat{x}$ ($\hat{y}$) direction as illustrated in Fig.~\ref{Fig:Fig6}(c). The single-valley network model does not have an analog of a trivial insulator because it is based on 1D chiral states, ``anomalous'' by construction.
Therefore, we take into account both valleys in the stripe-geometry calculation. We assume the two valleys, which are related by $\mathcal{C}_{2z}$ symmetry, to be decoupled in the bulk, but consider perfect inter-valley 180$^\circ$ reflections along edges where an incoming K-valley (K$'$-valley) state is {\it back} scattered at a junction [black dots in Fig.~\ref{Fig:Fig6}(c)] to an outgoing K$'$-valley (K-valley) state along the same domain-wall link. The boundary then separates the bulk from a trivial insulator. A detailed discussion can be found in Appendix~\ref{APP:OBC_cal}. In Fig.~\ref{Fig:Fig6}(b), we plot the energy spectrum of the stripe and confirm the existence of chiral edge states. Both the number and chirality of edge states are the same for every bulk gap, which agrees with the bulk topological invariant $\nu$. The existence of the edge state can be understood by the skipping orbits in the $P_{a\bar{b}}=1$ or $P_{a\bar{c}}=1$ limit, as shown in Fig.~\ref{Fig:Fig6}(c).
The chirality of edge states depends on $\sgn(\gamma-1)$, which is also consistent with the bulk butterfly diagram.

Therefore, our system realizes FTI, of which $n_{\text{edge}}$ is determined by the Floquet winding number instead of Chern numbers \cite{Kitagawa2010,Rudner2013,Pasek2014,Kim2020,Potter2020}. 
Here we  infer the topological invariant from the spectral flow of the Hofstadter butterfly, which provides a convenient way to diagnose the band topology. The invariant $\nu$ is also confirmed by explicitly computing the edge states.
Because of the periodic energy spectrum, we predict a cascade of chiral edge states with the same chirality and mode number prevailing in bulk gaps. We note that the energy spectrum of the network model describes the low-energy bands of mTBLG. The realistic mTBLG is a multi-band Chern insulator with the Chern bands submerged by other high-energy bands at the AB/BA regions. In addition, the topological properties can be modified by the energy dependence in the junction \cite{Potter2020}, which we ignore in this work.

\section{Discussion}\label{Sec:Discussion}

We have studied the electronic structure of mTBLG and calculated the Hofstadter butterfly of the network model in the presence of an out-of-plane magnetic field. We first construct a network model that describes a wide range of the scattering processes including the absence of $\mathcal{C}_{2z}\mathcal{T}$ symmetry.
Then, we show that Dirac points can reemerge at weak finite magnetic fields, providing fingerprints for the network model parameters. In the strong magnetic field regime, we demonstrate that the network model effectively realizes Floquet topological insulators \cite{Kitagawa2010,Rudner2013,Liang2013}, where Chern numbers of bulk bands are zero, but there are chiral edge states traversing bulk gaps.

Our predictions can be investigated by transport measurements in mTBLG. 
In mTBLG with $\theta=0.1^{\circ}$, the moir\'e period is $d \approx 140$ nm and the required magnetic field to generate one flux quantum per triangle is $B_0\approx0.49$T. 
It clearly suggests that the mTBLG is more advantageous in studying Hofstadter butterfly than graphene on hBN \cite{Dean2013,Ponomarenko2013} because the required magnetic field is even lower.

Our theory agrees with several existing experiments \cite{Rickhaus2018,Xu2019_GiantOscillation}, which all demonstrate that the domain-wall network dominates the transport at low temperature and for low density. 
For example, the magnetoresistance shows oscillations in $B$ which is consistent with the approximate periodicity in the magnetic flux \cite{Xu2019_GiantOscillation}. 
The sign alternation in the Hall resistivity implies the existence of Dirac points and van Hove singularities in the network model \cite{Efimkin2018,Xu2019_GiantOscillation}. 
The Dirac nodes can be identified by combining both the Hall resistivity (sign change) and the longitudinal resistivity (peaks). As we show in Fig.~\ref{Fig:Fig4}, the positions of Dirac nodes provide fingerprints to extract the parameters of the junction.
Therefore, our theory establishes a bridge between theory and experiment by studying the Dirac point positions in a weak magnetic field.

For the network model in a sufficiently large magnetic field so that ($\mathcal{C}_{2z} \mathcal{T}$ breaking) bulk gaps exceed disorder/thermal broadening, we predict an effective realization of the FTI in an \textit{equilibrium} system. 
In this topological state, chiral edge states with the same mode number and chirality can appear repeatedly in a cascade of low-energy bulk energy gaps.
Experimental observation of such a phenomenon would provide smoking-gun evidence for low-energy FTI. We note that the $\gamma$ as a function of the magnetic field is an important open question which is crucially related to the conditions of realizing FTI phase in the experiments.
Beside mTBLG, similar network models could also be realized in other Dirac systems with a periodic array of mass domains, for example, using the surface states of three-dimensional topological insulators.

We expect that the FTIs in our network model are robust against certain amount of disorder \cite{titum2016anomalous,Leykam2016}. The mTBLG features a new type of disorder, i.e., the twist angle disorder \cite{Wilson2019TBLGdisorder}, whose effect on the network model is an interesting future direction to pursue. 
In addition, many-body interaction can drive phase transitions \cite{Wu2019,Chou2019_SC_Int,Chen2019}. Electron repulsion can generate the FTIs in the absence of $B\hat{z}$ field by spontaneously breaking the $\mathcal{C}_{2z} \mathcal{T}$ symmetry, and may possibly lead to fractionalized states in the network model.

\section*{Acknowledgments} 

We thank Rui-Xing Zhang for stimulating discussions and Andrew Potter for comments on the Floquet topological insulator. This work is supported by the Laboratory for Physical Sciences. Y.-Z.C. is also supported by JQI-NSF-PFC (supported by NSF
grant PHY-1607611).

\appendix

\section{Network model under an out-of-plane magnetic field}\label{APP:B_field}

\begin{figure}[b]
	\includegraphics[width=0.4\textwidth]{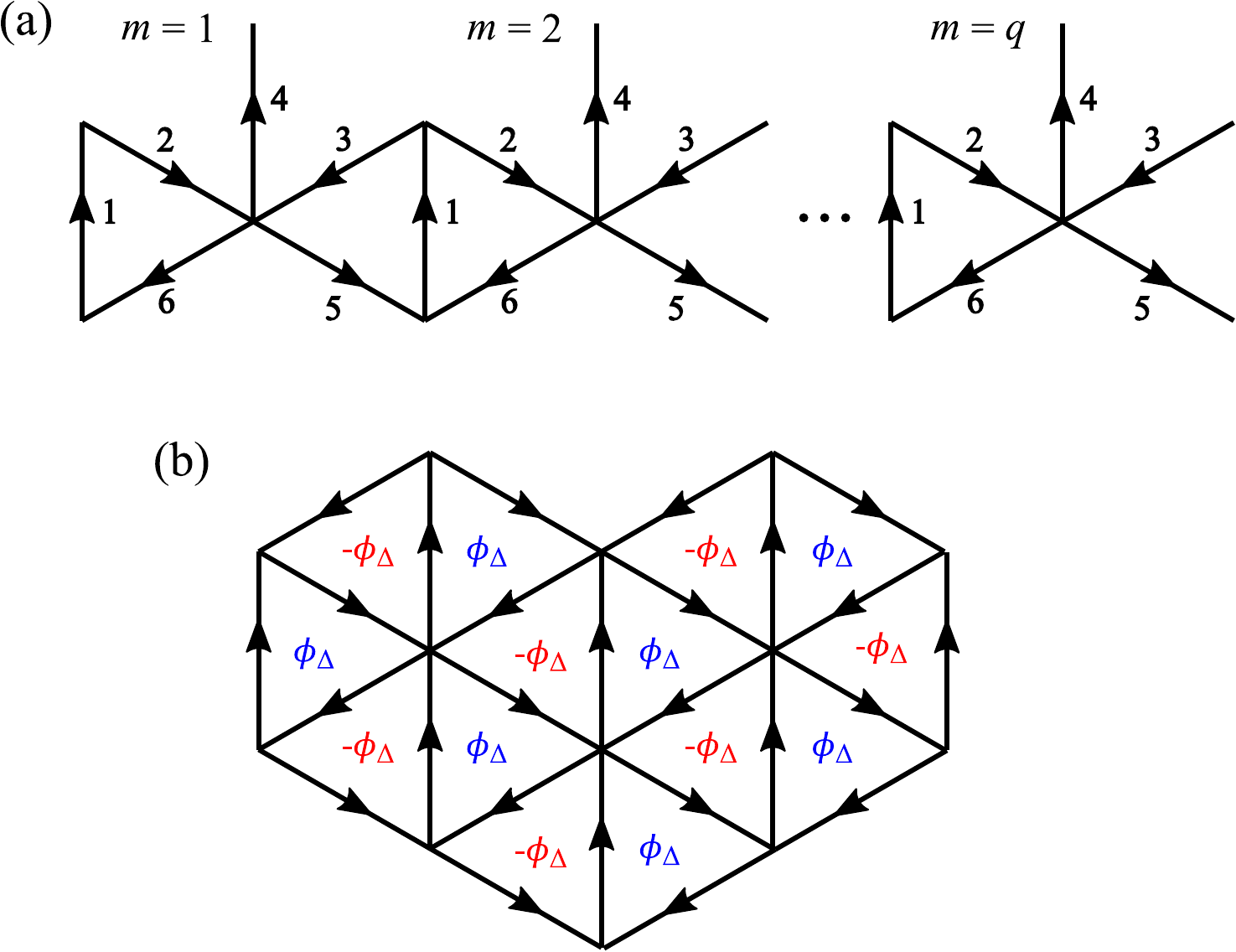}
	\caption{Magnetic unit cell and flux pattern in the network model. (a) Illustration of the magnetic unit cell. There are $m$ subcells that each contain 6 links. The phase accumulation for each link is given by Eq.~(\ref{Eq:phase_acc}). (b) The flux pattern induced by the magnetic field in the network model.}
	\label{Fig:S2}
\end{figure}

We discuss the procedures for calculating the energy spectrum in the presence of an out-of-plane magnetic field. We first choose the Landau gauge, $\vex{A}=(0,Bx)$, such that the translational symmetry is generically broken along the $\hat{x}$ direction. The commensurate limit ($eB\sqrt{3}d^2/h=p/q$ with $p$,$q$ being integers) is considered here. 
We choose a magnetic unit cell with and lattice vectors $\vex{\mathcal{R}}_1=(q\sqrt{3}d,0)$ and $\vex{\mathcal{R}}_2=(0,d)$ as illustrated in Fig.~\ref{Fig:S2}. 
The unit cell can be split into $q$ subcells that each contains 6 domain-wall links. The phase accumulations due to the Aharonov-Bohm effect (i.e., $- \frac{e}{\hbar} \int_\ell \vex{A} \cdot d \vex{r}$) are given by:
\begin{subequations}\label{Eq:phase_acc}
	\begin{align}
	\phi_{1,m}=&-\frac{2 (m-1)p\pi}{q}\\
	\phi_{2,m}=&\frac{p\pi}{4q}+\frac{(m-1)p\pi}{q},\\
	\phi_{3,m}=&\frac{3p\pi}{4q}+\frac{(m-1)p\pi}{q},\\
	\phi_{4,m}=&-\frac{p\pi}{q}-\frac{2p(m-1)\pi}{q},\\
	\phi_{5,m}=&\frac{3p\pi}{4q}+\frac{(m-1)p\pi}{q},\\
	\phi_{6,m}=&\frac{p\pi}{4q}+\frac{(m-1)p\pi}{q},
	\end{align}
\end{subequations}
where $\phi_{a,m}$ is the $a$th link in the $m$th subcell.

The flux pattern in the network model is presented in Fig.~\ref{Fig:S2}(b). Each clockwise (counterclockwise) loop encircles a flux $\phi_{\triangle}=\frac{p}{4q}$ ($-\phi_{\triangle}=-\frac{p}{4q}$).
With the magnetic unit cell introduced in Fig.~\ref{Fig:S2}(a), the spectrum can be straightforwardly calculated by imposing the Bloch equation. We generate butterfly diagrams by choosing all the prime numbers $q\le 37$ and $p=1,2,...,4q-1$. We consider a sufficiently fine grid in the Bloch wavevector space such that all the spectra converge. Practically, we need at least $N_x=4$ and $N_y=12$ where $N_x$ and $N_y$ are the number of wavevectors in $k_x$ and $k_y$ directions respectively.

Now we discuss the spectral flows in the butterfly diagrams, which reveal the topological properties of the band. 
We plot the special limits, $P_{a\bar{b}}=1$ and $P_{a\bar{c}}=1$ in Fig.~\ref{Fig:S3}. 
The spectral shift under the increase of a unit flux, $\delta\phi_{\triangle}=1$, is $-1/3$ ($1/3$) of $2\pi\hbar v/d$ for $P_{a\bar{b}}=1$ ($P_{a\bar{c}}=1$). Such energy shifts lead to $\nu=-1$ and $\nu=1$ for $P_{a\bar{b}}=1$ and $P_{a\bar{c}}=1$ respectively.
The amount of the spectral shift can be understood analytically in the following. For $P_{a\bar{b}}=1$, each electron move clockwise and encloses an isolated triangle. Each state thus contains three domain-wall links, and the magnetic flux encircled is exactly $\phi_{\triangle}$ [Fig.~\ref{Fig:S2}(b)]. Since each link picks up $\exp(i \frac{Ed}{\hbar v}-i \frac{e}{\hbar} \int_\ell \vex{A} \cdot d \vex{r} )$, we obtain $3\frac{\delta E d}{\hbar v}+2\pi\delta\phi_{\triangle}=0$ where $\delta E$ is the energy shift due to the change of $\phi_{\triangle}$. Similarly, we derive $3\frac{\delta E d}{\hbar v}-2\pi\delta\phi_{\triangle}=0$ for 
$P_{a\bar{c}}=1$.

\begin{figure*}[t]
	\includegraphics[width=0.9\textwidth]{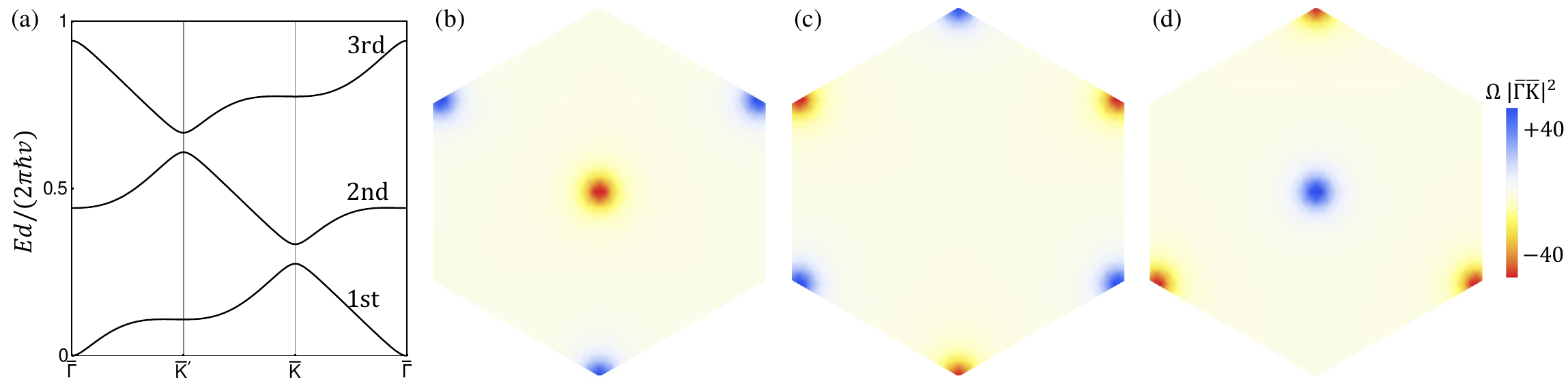}
	\caption{(a) Band structure of the network model in K valley. Parameters are  $P_{a\bar{a}}=1/9$ and $\gamma = 1.9$. Berry curvature $\Omega$ of the first, second and third bands in (a) are shown in (b), (c) and (d), respectively. Figures in (b)-(d) are plotted within the moir\'e Brillouin zone. $|\bar{\Gamma}\bar{K}|=4\pi/(3d)$ is the distance between $\bar{\Gamma}$ and $\bar{K}$ points.}
	\label{Fig:Berry}
\end{figure*}

Under the increase of one flux per triangle, the $n_0$th gap flows to the $n_1$the gap, where $\nu\equiv n_1-n_0$ is well defined. Electrons that follow this flow have a density $\delta\rho=\nu/\mathcal{A}_{\text{MUC}}$, where $\mathcal{A}_{\text{MUC}}= 2\mathcal{A}_{\triangle}$ is the area of one moir\'e unit cell (MUC) that encloses two triangles. Therefore, when the magnetic field increases by one flux per triangle, the number of electrons per triangle that follow the spectral flow is $\nu/2$. This might indicate that $\nu/2$ is the topological invariant. However, the above analysis only takes into account the K valley, whereas the full system includes both the K and K$'$ valleys.  Because the two valleys are related by $\mathcal{C}_{2z}$ symmetry, they have the same spectrum pattern with identical spectral flow. It is crucial that the two valleys must be considered together in order to get the correct topological invariant. The reason is that the network model  in each valley is based on 1D chiral states, and is therefore ``anomalous'' by construction. The topological invariant for the full system with both K and K$'$ valleys is given by $\nu$, which is also consistent with our edge state calculation.

\section{Berry Curvatures and Chern numbers}\label{APP:Chern_N}

We present an explicit calculation of Berry curvatures and Chern numbers for the network model. In the absence of an out-of-plane magnetic field, the $\mathcal{C}_{2z}\mathcal{T}$ symmetry enforces the Berry curvature to be identically zero. A finite Berry curvature is allowed once the  $\mathcal{C}_{2z}\mathcal{T}$ symmetry is broken (i.e., $\gamma \neq 1$). For simplicity, we consider a magnetic field that generates an integer flux $\phi_\triangle = N$, where $N \geq 1$ is an integer. The Aharonov-Bohm phases generated by this integer flux can be completely gauged away, and the moir\'e translational symmetry can be restored. In this case, the band energy and wave function are fully determined by the scattering $\hat{S}$ matrix with $\gamma \neq 1$:
\begin{equation}
\left(e^{-i\frac{E_{\vex{k},n} d}{\hbar v}}-\hat{T}^{-1}\hat{S}
\right) |u_{\vex{k},n}\rangle =0,
\end{equation}
where $\hat{T}=\text{diag}\left[e^{i\vex{k}\cdot\left(\vex{R}_1+\vex{R}_2\right)},e^{-i\vex{k}\cdot\vex{R}_2},e^{-i\vex{k}\cdot\vex{R}_1}\right]$ is the translation operator with $\vex{k}$ being the Bloch wave vector, and $\vex{R}_{1,2}$ are lattice vectors for the moir\'e lattice. $E_{\vex{k},n}$ and $|u_{\vex{k},n}\rangle$ are, respectively, the energy and wave function of the $n$th band at momentum $\vex{k}$.

We calculate the Berry curverature  $\Omega$ by numerically computing the Wilson loop. We discretize the momentum space by a rhombus grid, and each small grid spans a momentum-space area $\mathcal{A}_0=\mathcal{A}_{\text{MBZ}}/\mathcal{N}^2$, where $\mathcal{A}_{\text{MBZ}}$ is the momentum-space area of the moir\'e Brillouin zone (MBZ) and $\mathcal{N}$ is a sufficiently large integer characterizing the linear dimension of the grid. The Berry curvature is approximated by
\begin{align}
\nonumber&\Omega_n\left(\vex{k}=\frac{\vex{k}_1+\vex{k}_2+\vex{k}_3+\vex{k}_4}{4}\right)\\ 
\approx &\frac{\arg\left[ \langle u_{\vex{k}_1, n} | u_{\vex{k}_2, n} \rangle 
\langle u_{\vex{k}_2, n} | u_{\vex{k}_3, n} \rangle 
\langle u_{\vex{k}_3, n} | u_{\vex{k}_4, n} \rangle 
\langle u_{\vex{k}_4, n} | u_{\vex{k}_1, n} \rangle 
\right]}{\mathcal{A}_0},
\end{align}
where  $\vex{k}_1 \rightarrow \vex{k}_2 \rightarrow \vex{k}_3 \rightarrow \vex{k}_4 \rightarrow \vex{k}_1  $ tracks in a counterclockwise manner a small rhombus grid with the area $\mathcal{A}_0$. The Chern number $C$ of the $n$th band can then be calculated following the definition
\begin{equation}
C_n= \frac{1}{2\pi} \int_{\text{MBZ}} d \vex{k}  ~\Omega_n(\vex{k}).
\end{equation}

We show the calculated Berry curvature for a representative band structure of the network model with $\gamma \neq 1$ in Fig.~\ref{Fig:Berry}. For each band in Fig.~\ref{Fig:Berry}(a), the Berry curvature $\Omega$ reaches its maximum (positive) and minimum (negative) values around momentum points with {\it gapped} Dirac cones. The hot spots of $\Omega$ correspond to gap opening regions, as expected. The Chern number for each band shown in Fig.~\ref{Fig:Berry}(a) is identically zero, because the two hot spots of $\Omega$ in each band take opposite values and cancel out.

Therefore, we confirm through explicit calculation that the bands of the network model at an integer flux have a zero Chern number.

\section{Edge state and boundary condition}\label{APP:OBC_cal}

\begin{figure}[h]
	\includegraphics[width=0.4\textwidth]{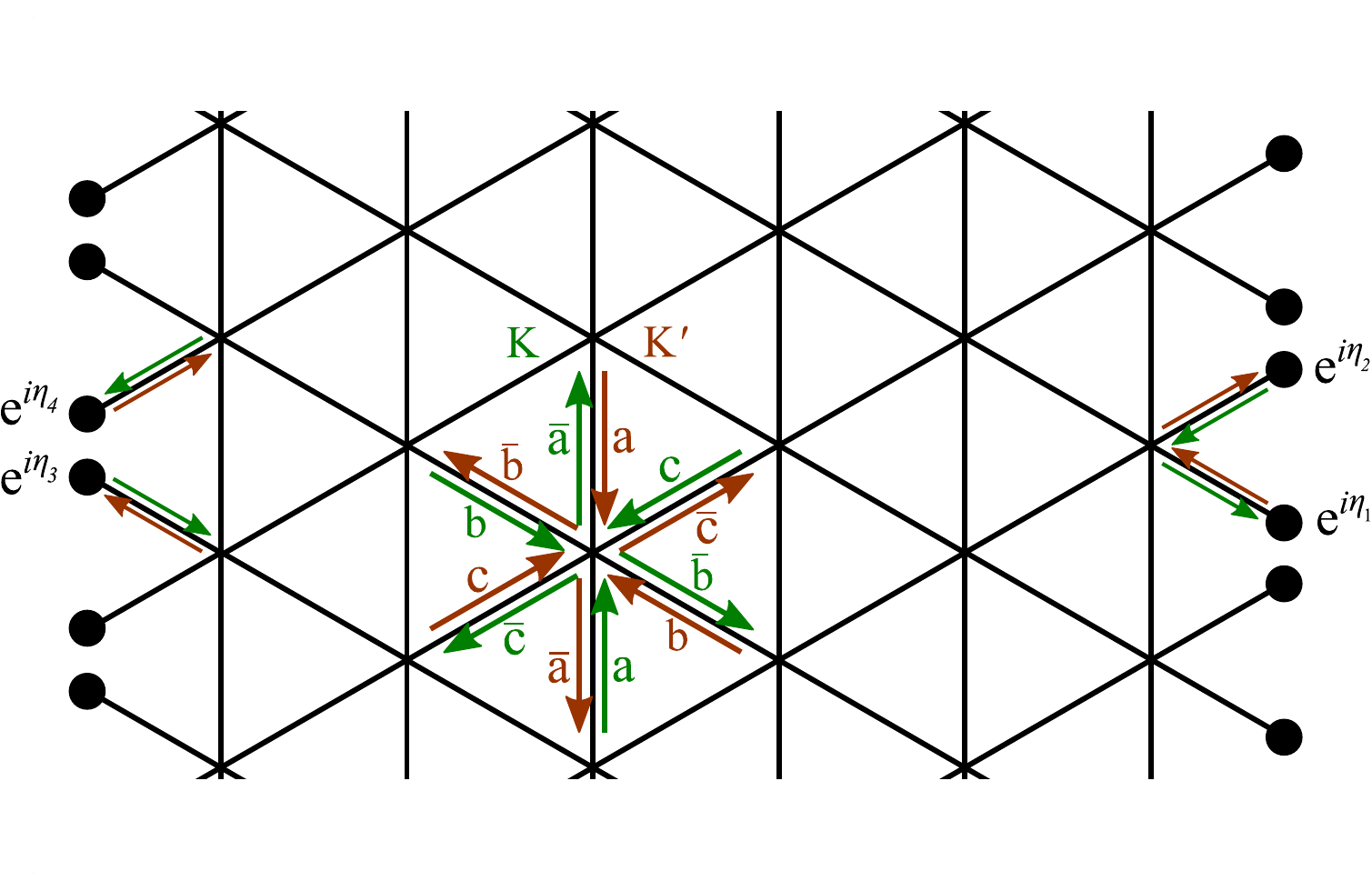}
	\caption{Network model with edges. We consider a strip with a finite width along $\hat{x}$ and a periodic boundary along $\hat{y}$. $\eta_1$, $\eta_2$, $\eta_3$, and $\eta_4$ are the reflection phases on the edges. }
	\label{Fig:S5}
\end{figure}

\begin{figure}[t!]
	\includegraphics[width=0.35\textwidth]{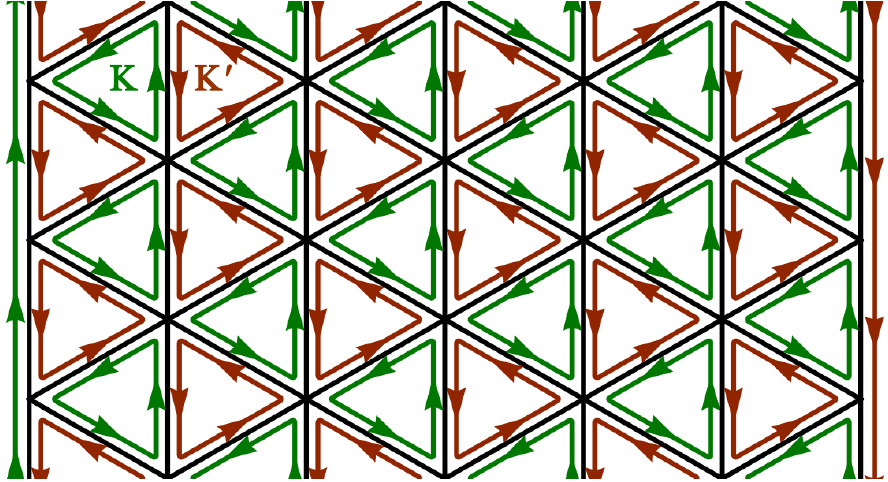}
	\caption{Edge state with a different boundary condition. The plot corresponds to $P_{a\bar{c}}=1$. Green arrows and brown arrows indicate K and K$'$ valley states respectively. The electron motions associated with the two valleys are related by $\mathcal{C}_{2z}$ symmetry (rotation center at AA region), and both are counterclockwise in this plot.}
	\label{Fig:S6}
\end{figure}

We provide detailed procedures for calculating the edge states in the network model. The main obstacle is that the single-channel network model with only the K valley is composed of the chiral states. One cannot construct a topologically trivial state from the chiral states because the chiral states are always anomalous. The resolution is to introduce both the K and K$'$ valleys such that there are two counter-propagating modes in each domain-wall link, i.e., non-chiral state. We will discuss the construction in depth in the rest of the section.

First of all, we digress and discuss how to construct a trivial state from a model including both the K and K$'$ valleys. We consider intervalley scatterings such that perfect 180$^{\circ}$ reflections take place at each junction. Incoming electrons in the K valley are scattered to outgoing electrons in the K$'$ valley and vice versa. 
Insulating states are therefore realized, and the wavefunction is localized within one segment.
This two-valley network model with perfect reflection realizes a topologically \textit{trivial} insulator.
The energy spectrum can be obtained analytically by solving the quantization condition $1=e^{i\frac{2Ed}{\hbar v}}e^{i(\eta+\eta')}$, where $\eta$ and $\eta'$ are the reflection phases at the end points. The energy spectrum is independent of momentum and the flatband energies are at $\hbar v [-(\eta+\eta')/2]/d$ and $\hbar v [\pi-(\eta+\eta')/2]/d$ within one energy period.

Now, we discuss how to study the two-valley network model with boundaries. The setup is illustrated in Fig.~\ref{Fig:S5}.
We consider both K and K$'$ valleys, which are connected by a $\mathcal{C}_{2z}$ operation in the bulk, where scattering at the junctions is characterized by a matrix as follows:
\begin{align}
\left[\begin{array}{c}
\hat\Psi_{\text{out}}^{\text{K}}(\vex{x}_0)\\[2mm]
\hat\Psi_{\text{out}}^{\text{K}'}(\vex{x}_0)
\end{array}
\right]=
\left[\begin{array}{cc}
\hat{S} & 0\\[2mm]
0 & \hat{S}
\end{array}
\right]
\left[\begin{array}{c}
\hat\Psi_{\text{in}}^{\text{K}}(\vex{x}_0)\\[2mm]
\hat\Psi_{\text{in}}^{\text{K}'}(\vex{x}_0)
\end{array}
\right],
\end{align}
where $\hat{S}$ is given by Eq.~(1) in the main text,
\begin{align}
\hat\Psi_{\text{in}}^{\text{K}/\text{K}'}=\left[\begin{array}{c}
\psi_a^{\text{K}/\text{K}'}\\
\psi_b^{\text{K}/\text{K}'}\\
\psi_c^{\text{K}/\text{K}'}
\end{array}\right],\,\,\,
\hat\Psi_{\text{out}}^{\text{K}/\text{K}'}=\left[\begin{array}{c}
\psi_{\bar{a}}^{\text{K}/\text{K}'}\\
\psi_{\bar b}^{\text{K}/\text{K}'}\\
\psi_{\bar c}^{\text{K}/\text{K}'}
\end{array}\right].
\end{align}
The intervalley scattering in the bulk is completely ignored. 
In addition, the scattering matrix is the same for both K and K$'$ microscopic valleys because of $\mathcal{C}_{2z}$ symmetry. It is worthwhile to note again that the 1D domain-wall states associated with the two valleys have opposite chiralities, as also required by $\mathcal{C}_{2z}$ symmetry.

Finally, we consider scattering on the boundaries. The perfect reflecting boundary conditions are imposed on the edge, which can be viewed as the interface between the bulk network model and topologically trivial insulators. The black dots in Fig.~\ref{Fig:S5} indicate those perfect reflecting junctions. The scatterings at the rightmost junctions are given by
\begin{align}
\psi_{\bar{b}}^{\text{K}'}=e^{i\eta_1}\psi_{{b}}^{\text{K}},\,\,\,\psi_{\bar{c}}^{\text{K}}=e^{i\eta_2}\psi_{{c}}^{\text{K}'};
\end{align}
similarly, the scatterings at the leftmost junctions are given by
\begin{align}
\psi_{\bar{b}}^{\text{K}}=e^{i\eta_3}\psi_{{b}}^{\text{K}'},\,\,\,\psi_{\bar{c}}^{\text{K}'}=e^{i\eta_4}\psi_{{c}}^{\text{K}}.
\end{align}
In the main text, we choose $\eta_1=\eta_2=\eta_3=\eta_4=0$. The topological properties are independent of the choice of the reflection phases.

The existence of the edge states can be understood by the skipping orbits in a special limit, $P_{a\bar{c}}=1$ (or equivalently $P_{a\bar{b}}=1$). In Fig.~\ref{Fig:S6}, we provide an alternative way to construct the edge state. The system is also in a strip geometry with a finite width in $\hat{x}$ and a periodic boundary condition in $\hat{y}$. The electrons in the bulk move counterclockwise in localized triangular loops. On the other hand, the right and left boundaries host perfect forward moving edge states because each domain-wall link must contain a pair of counter propagating movers.

%%\bibliography{Butterfly}

%%%%%%%%%%%%

%%%%%%%%%%%%

\end{document}